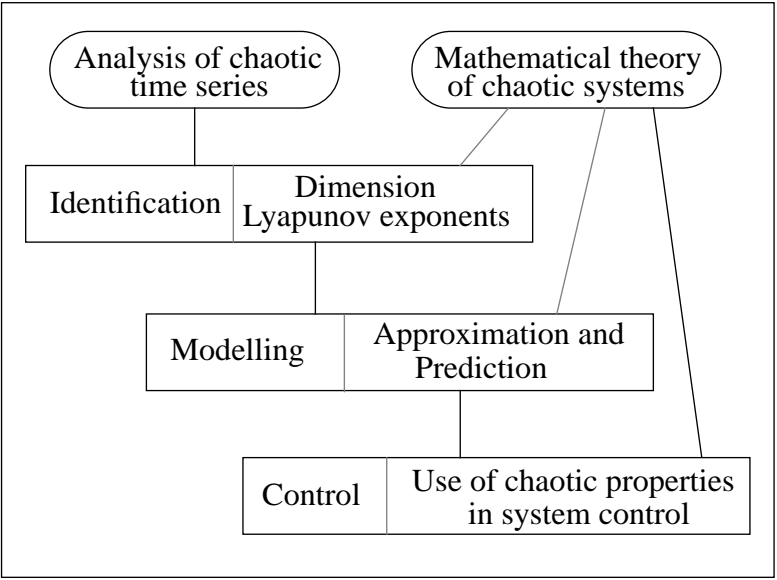

# Chaotic time series
# Part I: Estimation of invariant properies in state space

D. Kugiumtzis [*], B. Lillekjendlie [†], N. Christophersen [†]

January 14, 1994

**Abstract**

Certain deterministic non-linear systems may show chaotic behaviour. Time series derived from such systems seem stochastic when analyzed with linear techniques. However, uncovering the deterministic structure is important because it allows for construction of more realistic and better models and thus improved predictive capabilities. This paper describes key features of chaotic systems including strange attractors and Lyapunov exponents. The emphasis is on state space reconstruction techniques that are used to estimate these properties, given scalar observations. Data generated from equations known to display chaotic behaviour are used for illustration. A compilation of applications to real data from widely different fields is given. If chaos is found to be present, one may proceed to build non-linear models, which is the topic of the second paper in this series.

## 1 Introduction

Chaotic behaviour in deterministic dynamical systems is an intrinsicly non-linear phenomenon. A characteristic feature of chaotic systems is an extreme sensitivity to changes in initial conditions while the dynamics, at least for so-called dissipative systems, is still constrained to a finite region of state space called a strange attractor. Time traces of the state variables of such systems display a seemingly stochastic behaviour.

The systematic study of chaos is of recent date, originating in the 1960s. One important reason for this is that linear techniques, so long dominant within applied mathematics and the natural sciences, are inadequate when considering chaotic phenomena. Furthermore, computers are a necessary tool for studying such systems. As a result, the amazingly irregular behaviour of some non-linear deterministic systems was not appreciated and when such behaviour was manifest in observations, it was typically explained as stochastic. Although the lasting role of chaos in the natural sciences and other disciplines cannot be assessed at present, it seems clear that the techniques being developed within this rapidly evolving field offer additional methods for analysis and modelling of dynamical systems which deserve a place alongside more traditional techniques. In this connection it is worthwhile to note that many types of non-linear equations may give rise to chaotic behaviour. Thus if one is interested in non-linear systems but not chaos *per se*, the model or system under study may still be chaotic in parts of the parameter space. To diagnose and understand (or prevent) such situations, knowledge of chaos is necessary.

Two separate, but interacting lines of development characterize chaos theory. A theoretical line focuses on systems of ordinary non-linear difference and differential equations that may show chaotic behavior. Stephen Smale initiated the modern work here in 1963 by constructing a simple mapping embodying the salient featues of chaotic systems [81], [82]. This is largely a mathematical discipline involving concepts such as bifurcations and stable and unstable manifolds [43], [97], [4]. A more applied direction stems from the meteorologist Edward Lorenz, who, also in 1963, observed extreme sensitivity to changes in initial conditions of a simple non-linear model simulating atmospheric convection [60]. This more experimental approach, which is the topic of our review, relies heavily on the computational study of chaotic systems and includes methods for investigating potential chaotic behaviour in observational time series [66] [93], [79] and [39].

[*]Department of Informatics, University of Oslo, Pb. 1080 Blindern, N-0316 Oslo, Norway
[†]SINTEF-SI, Pb. 124, Blindern, N-0314 Oslo, Norway

Chaos has been identified in simple experiments such as a water dripping faucet [18], simple electric circuits [10], and in situations involving near turbulent flow such as the Couette-Taylor experiment [9]. Outside the laboratory, chaotic behaviour has been claimed in climatic time series, [70], [26], [36], astrophysics [46], hydrodynamics [47], economics [12], medicine [5] and several other fields. Clearly one is still in an exploratory phase where chaos is a hot topic and such behaviour is sought in many diverse areas. It is generally too early to judge the final contribution of chaos theory in many of these fields.

The theory of *fractals* has evolved in parallel with chaos theory and these two fields have an important linkage. Fractals, brought to the forefront by Benoit Mandelbrot [62], are objects that are self similar on different scales. They have become popular partly due to the computer generated pictures that have been produced, but fractal mathematical objects such as Cantor and Julia sets have been known for a long time. In chaos theory, the so-called *strange attractors* are objects with fractal properties. These are the geometrical stuctures traced out in state space by the trajectories of chaotic systems.

As noted above, time series derived from deterministic chaotic systems appear to be stochastic, but standard linear modelling and prediction techniques, such as ARMA models, are not suitable for these systems. For instance, a simple scalar, non-linear system known as the logistic map gives rise to time series with the same autocorrelation function as white noise. It is therefore natural to ask what is the difference between a chaotic system and a stochastic one. We will not discuss this rather deep question in any detail (c.f. [57]), but take a pragmatic approach and characterize the systems that lend themselves to analysis by the techniques presented below as predominantly chaotic. More technically, this implies low dimensional systems with positive Lyapunov exponents.

The relatively new methodology described in this and the accompanying paper is based on concepts from mathematical topology and dynamical systems theory, as well as information and numerical approximation theory. Knowledge related to the fields from which data originate is not involved in this analysis, but this is clearly necessary when interpreting the results.

The study of potentially chaotic systems may be divided into three areas: *identification of chaotic behaviour*, *modelling and prediction*, and *control* (see Fig(1)). The first area, covered in this paper, is important because it shows how chaotic systems may be separated from stochastic ones and, at the same time, provides estimates of the degrees of freedom and the complexity of the underlying chaotic system. Based on such results, identification of a state space representation allowing for subsequent predictions may be carried out. These topics are reviewed in the accompanying paper. The last stage, if desirable, involves control of a chaotic system. In this area, one may actually use the chaotic behaviour to advantage, obtaining a "large" desired effect at the expense of a "small" control signal [80], [20]. One may even speculate that non-linearities, making a system chaotic, could be introduced for control purposes to take advantage of such behaviour. This is a rather new research area, strongly involving the mathematical and theoretical aspects of chaos theory, and is not covered by our review [1].

Figure 1: A sketch of links between areas of chaotic system analysis and synthesis.

Following this introduction, some basic features of chaotic systems are summarized in Section 2; in Section 3, the problem of reconstructing state variables from time series is discussed. In the last two sections, two basic characteristics of chaotic dynamics are treated, namely the dimension of the strange attractor and the Lyapunov exponents.

# 2  Background

Here we give a short description of chaotic systems assuming that the state equations are known. Such systems exhibit unpredictable behavior in the sense that for given initial conditions with finite precision, the long term behaviour cannot be predicted, except to say that the states are constrained to a certain finite region of state space. Initially nearby trajectories may divergence exponentially for chaotic systems. Still, short time predictions are feasible. Weather forecasting offers a parallel here. Climatic data show a limited variability, but forecasts are impossible beyond 10-14 days, no matter the amount of available data.

Assume that the dynamical equations are given for continuous time as $\dot{\underline{s}}(t) = f(\underline{s}(t))$ and for discrete time as $\underline{s}_{k+1} = f(\underline{s}_k)$, where $\underline{s}$ is the state vector of dimension $n$, $f$ is the system function,[1] and $t$ and $k$ are the continuous and discrete time variables, respectively. These systems are autonomous and completely deterministic.

Chaotic systems may have a very simple form as for example the logistic map $s_{k+1} = a s_k (1 - s_k)$. For $a$ in the range 1 - 4, the state variable is restricted to the interval [0,1] for initial conditions in the same interval. Increasing the value of $a$ from 1 to 4 changes the asymptotic state from a fixed point through stable periodic orbits with increasing periodicity, to chaotic behaviour at $a \geq 3.57$ [44]. The logistic map has been used to model populations in biology [65] and supply and demand in economics [52]. A realization of the logistic map for $a = 4$ is given in Fig(2) where successive points in time have been connected by straight lines. The autocorrelation function for this sequence is given in Fig(3). It is essentially non-zero only at the origin, and a natural conclusion would have been that we have white noise. The logistic map in the chaotic domain was used as a random number generator in early computers [94].

Figure 2: The logistic map in a chaotic state ($a = 4$).

Figure 3: The autocorrelation function for the logistic map in a chaotic state ($a = 4$).

A strange attractor is the geometrical structure formed by the asymptotic states of a chaotic system. On this attractor, the dynamics are characterized by *stretching* and *folding*; the former phenomenon causes the divergence of nearby trajectories and the latter constrain the dynamics to a finite region in a subspace of dimension $\leq n$. This subspace is the smallest space *embedding* the attractor. In contrast to non-chaotic systems having attractors of integer dimensions (e.g. points and limit cycles), chaotic systems have strange attractors characterized by a non-integer dimension $d$. (Various definitions of non-integer dimensions are given in Section 4.) This dimension is a property of the dynamical system, independent of any particular trajectory. Such properties are called **invariants** of the system. As an example, consider the famous Lorenz system: [60] [2]

$$\begin{aligned}
\dot{s}_1 &= -a(s_1 - s_2) \\
\dot{s}_2 &= -s_1 s_3 + b s_1 - s_2 \\
\dot{s}_3 &= s_1 s_2 - c s_3
\end{aligned} \qquad (1)$$

For suitable values of the parameters $a$, $b$ and $c$ this system is chaotic (Fig(4)). Despite the flat appearance of the attractor, it has a so-called correlation dimension of 2.06 and the smallest embedding dimension is 3, the same as the number of state variables.

Figure 4: The Lorenz attractor for the parameter values $a = 10$, $b = 28$ and $c = 8/3$.

---

[1] For simplicity, we use the same notation for the system function in the continuous and discrete case. Usually one assumes $f$ to be continuously differentiable.

[2] These equations arise in the study of thermal convection in the atmosphere and in the theory of high-powered lasers .

Another well known example with different properties is the Mackey-Glass delay differential equation:

$$\dot{s} = \frac{0.2s(t-\Delta)}{1+[s(t-\Delta)]^{10}} - 0.1s(t), \tag{2}$$

where $\Delta > 0$ is a delay parameter. This equation is a model for generation of blood cells in patients with leukemia [61]. Delay differential equations as well as partial differential equations are infinite dimensional, i.e. an infinite set of numbers is required to specify the initial conditions. For initial values of $s$ in the time interval $[0, \Delta]$, the system approaches a stable equilibrium point for $\Delta < 4.53$, a limit cycle for $\Delta \in [4.53, 13.3]$, and after a series of period doublings for $\Delta \in [13.3, 16.8]$ the system becomes chaotic. The strange attractor has dimension 1.95, 2.44, 3.15 and 7.5 for $\Delta$ equal to 17, 23, 30, and 100, respectively [27]. Note that dissipative partial differential equations may also have attractors of finite dimension [88].

Another invariant characterizing chaotic systems is the set of Lyapunov exponents which measures the average local degrees of expansion and contraction (see Fig(5)). The Lyapunov exponents are related to the eigenvalues of the linearized dynamics across the attractor. Negative values show stable behaviour while positive values show unstable behaviour. For chaotic systems, being both stable and unstable, the Lyapunov exponents indicate the complexity of the dynamics. The largest positive value determines the upper prediction limit. High dimensional chaotic systems tend to have very large positive exponents and predictions may be of little use. We will focus on low dimensional systems defined somewhat arbitrarily as having an attractor dimension $\leq 10$.

Figure 5: The expansion and contraction of the Lorenz system as characterized by Lyapunov exponents on a sphere centered on the trajectory shown. Here, the Lyapunov exponents are: $\lambda_1 = 1.5$, $\lambda_2 = 0$ and $\lambda_3 = -22.5$.

## 3 Reconstruction of state space

We assume discrete scalar observations of the chaotic system in the sequel, which is the case covered by the present theory. Let us denote the time series by $\{x(k\tau_s)\}=\{x_k\}$, where $k = 0, ..., N$ and $\tau_s$ is either the sampling interval (for continuous systems) or the discrete time interval. The time series is related to the state vector of the underlying dynamical system by $x_k = h(\underline{s}(k\tau_s))$, where $h$ is the measurement function. Apart from the measurements and the sampling time, the system function $f$, the attracor dimension $d$, and the measurement function $h$ are unknown.

Since the dynamics are unknown, we cannot reconstruct the original attractor that gave rise to the observed time series. Instead, we seek an **embedding space** where we can reconstruct an attractor from the scalar data that preserves the invariant characteristics of the original unknown attractor. The embedding dimension $m$ of the reconstructed state space will in general be different from the unknown dimension $[d]+1$, where $[d]$ denotes the integer part of $d$ (the attractor's dimension). The simplest method of deriving a state vector $\underline{x}_k$ is that of **delay coordinates** proposed by [73]:

$$\underline{x}_k = [x_{k\tau}, x_{k-\tau}, \ldots, x_{k-(m-1)\tau}]^T \tag{3}$$

$k = m-1, m, ....., N$. Here $\tau$ is an integer multiple of $\tau_s$ called the **time delay**. The introduction of $\tau$ allows for the possibility of skipping samples during the reconstruction. According to this method, the reconstruction of the state vector simply consists of determining the embedding dimension $m$ and the time delay $\tau$. Other methods are also possible, for example assigning state variables to successively higher order derivatives of $x(t)$ for continuous systems. However, this method is impractical due to noise amplification.

The above approach to reconstructing the attractor is based on *Taken's theorem* which guarantees that for an *infinite noise free data series* one can almost always find an embedding dimension $m$ preserving the invariant measures [87], [64]. Takens proved that under these conditions it is sufficient that $m \geq 2d + 1$ [3]. Under the idealized conditions of the theorem, the actual values of $\tau_s$ and $\tau$ are irrelevant; in practice, however, this is not the case. The theorem guarantees that the attractor embedded in the $m$-dimensional state space is "unfolded" without any self intersections. The condition $m \geq 2d + 1$ is sufficient but not necessary, and an attractor may be reconstructed successfully, also in practice, with an embedding dimension as low as $[d] + 1$. As an example let us consider the Henon map:

$$s_1(k+1) = s_2(k) + 1 - 1.4s_1(k)^2$$
$$s_2(k+1) = 0.3s_1(k)$$
(4)

This system is chaotic and has an attractor of dimension 1.26 shown in Fig(6(a))(see for example [3]). In Fig(6(b)), we reconstructed the attractor from single measurements of $x = s_1$ using embedding dimension 2 and $x(k)$, $x(k+2)$ as state space coordinates. One notices some self intersections of the reconstructed attractor of Fig(6(b)) that disappear with $m = 3$ as shown in Fig(6(c)). According to Taken's theorem $m = 4$ would be sufficient to guarantee reconstruction. If we choose coordinates $x(k)$, $x(k+1)$, the attractor is successfully reconstructed even with $m = 2$. However, for another choice of coordinates one may need to go as high as $m = 4$.

Figure 6: Reconstruction of the Henon attractor. The original attractor in (a); the reconstructed attractor from measurements of $s_1$ with $\tau = 2$ and $m = 2$ in (b) and with $\tau = 2$ and $m = 3$ in (c).

In reality, with a finite number of noisy data, the estimates of the invariant measures $d$ and the Lyapunov exponents are found to depend on both $m$ and $\tau$. Thus the central task of delay reconstruction is to choose these values properly. Various methods have been suggested but decisions are still often subjective in the end and related to the problem at hand.

## 3.1 Choosing Embedding Dimension

One approach to this problem is the so-called **the singular system approach** [14], [4] widely used in many areas of applied numerical linear algebra. The method assumes an initial reconstruction of the state space with an arbitrary large dimension $m$, even larger than that suggested by Taken's theorem. From the sequence of state vectors $\underline{x}_k$ one forms the $m \times m$ matrix $V = X^T X$ where the rows of $X$ are the state (row) vectors $\underline{x}_{m-1}$, $\underline{x}_m$ etc. or some scaled version of these vectors. The eigenvalues of $V$ are computed, typically using the well known singular value decomposition method present in many subroutine libraries. The eigenvectors of $V$ corresponding to the largest eigenvalues, are the directions in the reconstructed state space showing the largest variations in the data. This analysis can also be carried out locally for a neighborhood of a given data point in state space instead of the whole ensemble [13].

Assuming that the dominant variation in the data is due to the chaotic dynamics, the user will select $m$ equal to the number of "large" eigenvalues of $V$. The method has the advantage of reducing the effects of noise, especially white noise. However, the user has to decide subjectively the number of eigenvalues to retain and this may not be obvious in practice. Furthermore, the relative magnitude of the eigenvalues will depend on which variant of the $V$ matrix one factors. Other drawbacks of the method are that the resultant reconstruction is not always optimal and that it fails to distinguish a chaotic signal from noise with nearly the same Fourier spectrum [33].

---

[3] Others refer to it as Whitney's theorem because he first proved that a smooth $j$-dimensional manifold may be embedded in $\Re^{2j+1}$ [96].

[4] This method is referred to in the literature under many names including *principal component analysis*, *factor analysis* and *Karhunen-Loeve decomposition*

Another method for estimating $m$ is to study the geometrical structure of the attractor while embedded in successively higher dimensions. If $m$ is too low, the attractor displays self intersections; spatially nearby points on the attractor (but not necessarily temporally close) are either real neighbors due to the system dynamics or false neighbors due to the self intersections. In a higher dimension, where the self intersections are resolved, the false neighbors are revealed as they become more remote. One tries to find a threshold value for $m$ where no false neighbors are identified as one moves to increased dimensions. Two independent workers have put this approach into algorithmic form, [58] and [54]. The latter work explicitly considered noisy data and came with the expected conclusion that noise increases the estimated embedding dimension. Another related geometrical approach has also recently been proposed but instead of working with points on the attractor, the tangent of the trajectory is introduced [53]. In part 2, where identification and prediction are discussed, embedding dimensions are estimated as part of some methods.

## 3.2 Choosing Delay Coordinates

For too small $\tau$, the coordinates of $\underline{x}_k$ in the reconstructed state space are fairly similar and the attractor is stretched along a diagonal and easily obscured by noise. A choice of $\tau$ that keeps the coordinates more time independent is desirable. On the other hand, too large values cause loss of information contained in the data and two temporally close vectors become rather remote, giving rise to uncertainties.

A common, but ad hoc, choice for $\tau$ is the time at which the autocorrelation function has its first zero. Then the coordinates are *linearly uncorrelated*. A more sophisticated choice for $\tau$ has been suggested by Fraser and Swinney [34] giving a criterion of more general independence measured as the information in bits gained about $x(t+\tau)$ given the measurement of $x(t)$. On the average, this is the **mutual information** $I(\tau)$ and its first minimum is suggested as a good estimate for $\tau$. Still there is no guarantee that the mutual information has a clearcut minimum, given the time resolution $\tau_s$.

## 4 Dimensions of a strange attractor

There has been a strong interest in research on dimensions of chaotic attractors which has been paralleled by similar work on fractals. Even before 1980, many different theoretical definitions of dimension had been proposed, all based on scaling laws. In the beginning of the 1980's, tractable algorithms were derived and they were applied in widely different fields. The results seemed to support chaotic behaviour in many cases as low dimensional estimates were found. Today, these algorithms are known to have limitations and drawbacks, and caution seems warranted. For a comprehensive review of the topic, we refer to articles by Eckmann and Ruelle [22] and Theiler [90].

### 4.1 Measures of Dimension

The first definition of a non integer dimension was given in 1919 by Hausdorff [48] but its abstract form was not suitable for practical computation. Nevertheless, it has formed the basis for other definitions and subsequent algorithms. Closely related to the Hausdorff dimension is the **fractal dimension**, based on scaling the mass of the attractor with size. It is also referred to as the box counting dimension due to the way it is computed [30] (for the first numerical algorithm see [77]). Let $M(l)$ be the number of hypercubes of a given dimension $m$ with side length $l$ required to cover the attractor. Then from the scaling law

$$M(l) \sim l^{-D} \qquad (5)$$

the fractal dimension $D$ is derived

$$D = \lim_{l \to 0} \frac{\log[1/M(l)]}{\log l}. \qquad (6)$$

Using this definition, the dimensions of a point, a line, and an area in two dimensional space are the usual values 0, 1, and 2, respectively. To see this, take a square of side $l$ and note that the number needed to cover a point is proportional to $1/l^0$, to cover the line $M(l) \sim l^{-1}$, and to cover a surface $M(l) \sim l^{-2}$.

The definition has a number of practical limitations [42]; however, there is an efficient algorithm for the estimation of $D$ in [59].

In the definition of the fractal dimension, only the geometrical structure of the attractor is taken into account without considering the distribution of points on the attractor; all cubes count the same even though the frequencies with which they are visited may be very different. This is taken into account by the so-called **information dimension** [99] [28] which is derived from the minimal information needed to specify a point in a set, such as a hypercube, to an accuracy $l$. This information is measured as

$$S(l) = -\sum_{i=1}^{M(l)} p_i \log p_i, \qquad (7)$$

Here $p_i$ is the probability of a point being in the $i$th set defined as $p_i = \mu_i/N$ where $N \to \infty$ and $\mu_i$ is the number of points in the $i$th set. The information dimension $\sigma$ of the attractor is then given by

$$\sigma = \lim_{l \to 0} \frac{-S(l)}{\log l}. \qquad (8)$$

The most popular measure of an attractor's dimension is the **correlation dimension**, first defined by Grassberger and Procaccia [40]. It can be considered a simplification of the information dimension:

$$\nu = \lim_{l \to 0} \frac{\log \langle \mu_i \rangle_x}{\log l}, \qquad (9)$$

where $\mu_i$ is defined as before for a ball with center $\underline{x}_k$ instead of a cube and $\langle . \rangle_x$ denotes the average over all points $\underline{x}_k$.

The general rule $\nu \leq \sigma \leq D$ holds for the three different measures with equality when the points are distributed uniformally over the attractor [50], [41]. A fourth dimension is related to the Lyapunov exponents (see below). However, all measures are generally similar, and we will use the symbol $d$ to denote any of them.

In practice, the distance $l$ cannot approach zero and the usefulness of these measures depend on the corresponding computational algorithms. We will focus on the correlation dimension because it has the best computational properties and has been applied extensively in the last decade. However, part of the discussion is valid for the other measures as well.

## 4.2 The correlation dimension

The computation of the correlation dimension, often referred to as the Grassberger-Procaccia (GP) method, starts by estimating $\langle \mu_i \rangle_x$ using

$$C(l) = \frac{1}{N(N-1)} \sum_{i,j=1}^{N} \theta(l - |\underline{x}_i - \underline{x}_j|), \qquad (10)$$

where $\theta(x)$ is the Heaviside step function defined as $\theta(\rho) = 1$ for $\rho \geq 1$ and $\theta(\rho) = 0$ for $\rho < 0$. (The quantity $C(l)$ for $\lim_{N \to \infty}$ is termed the *correlation integral*.) The scaling of $C(l)$ with distance $l$ gives the estimate $\nu$ of the attractor's dimension, $C(l) \propto l^\nu$.

When the embedding dimension is unknown, $C(l)$ is computed for different values of $m$. If the system is chaotic, the slope of $\log C(l)$ versus $\log l$ converges to $\nu$ over an appropriate interval as $m$ increases. The rationale behind this approach follows from the discussion of the unfolding of the attractor. When the dimension of the working space is too low, the data will fill the entire accessible state space and the slope of the graph is equal to $m$. As $m$ approaches the limit in Taken's theorem, one expects the slope to approach $d$ and remain at that value. Actually, the embedding dimension for which the slope first seems to have converged, can serve as an estimate for the embedding dimension needed to reconstruct the state space.

Fig(7) shows this situation for the Lorenz system. Note the convergence of the $\log(C(l))$ for increasing embedding dimensions to a line of constant slope (of approximately 2.06) over a sizable interval of $\log l$.

Figure 7: Graphs of the correlation integral for embedding dimensions $2, \ldots, 10$ derived from 10000 measurements of the $s_1$-variable in the Lorenz system using $\tau = 15$ (the first minimum of the mutual information).

The computational demands scale as $O(N^2)$ for simple versions of the GP algorithm, and there have been many suggestions for improving the computational efficiency. For instance, it is unnessecary to compute interpoint distances that are larger than the current value of $l$. To achieve this, Theiler [89] and Grassberger [38] propose to organize the data in hypercubes, while Bingham [8] suggests using multidimensional trees. Other algorithms allow simultaneous computation of $C(l)$ for a range of embedding dimensions [31]. With such techniques one can approach computational demands proportional to $N \log N$ depending on the values of the parameters involved.

## 4.3 Practical considerations for dimension estimates

When working with simulated data from known equations taken over an extended period as in Fig(7), the scaling law is observed over an appreciable interval of $\log l$. But deviations occur in the upper and lower ranges. In the upper range, all graphs converge to 0 as $l$ aproaches the diameter of the attractor, while in the lower range the scaling law breakes down as $l$ approaches the minimum distance between points in the embedding space. Thus three separate regions can be identified in the diagram.

With real observations, one hopes to observe the scaling law for an intermediate $\log l$ interval of reasonable length. However, subject to the quality and size of the data as well as to the complexity of the underlying chaotic system, the scaling region is often short and sometimes not even discernable. The correlation dimension algorithms are robust to low-amplitude noise; in this case the scaling brakes down at distances around the noise amplitude and a fourth region is observed at low to intermediate values of $\log l$ [9] (see Fig(8)).

Figure 8: The four typical intervals of $\log l$ in the estimation of the correlation dimension from noisy data.

Certain modifications of the GP algorithm may enhance the performance in the case of higher amplitude noise [21], [85]. Linear low-pass filtering in this connection is not recommended because it may increase the estimated dimension [6], [68], but non-linear filtering has been successfully applied [55], [56], [45], [29].

Concerning noise, a natural queston is if correlation dimension algorithms may be used to separate between predominantly chaotic signals and stochastic ones. When the time series is white noise, for instance, one expects that the estimated dimension always coincides with the embedding dimension because the reconstructed attractor will always fill the embedded space, no matter its dimension. This is a powerful tool for distinguishing chaotic time series from white noise; however, it is more complicated to separate chaos from correlated noise. One approach is to compare the original time series with an artificial stochastic series with the same power spectrum, typically by taking a Fourier transform of the original time series, randomizing the phase, and then inverting back to the time domain. If the correlation integrals derived from the two data sets scale significantly different, the original time series cannot be stochastic ([37], [90]). The importance of this test is reinforced by the striking results in several publications over the last four years demonstrating that some types of correlated time series (i.e. stochastic time series with $1/f^\alpha$ power spectrum) exhibit low correlation dimensions ([71], [92], [75]).

In some cases, the attractor may have properties that complicate the computation of the correlation dimension. The so-called *lacunarity*, first introduced by Mandelbrot [63], is such a property and causes oscillations in the graph of $\log C(l)$ [84]. Some attractors are characterized by the *edge effect*, i.e. the neighbor points on the edge of the attractor follow a different scaling than the points in the interior. An attempt to circumvent this effect is discussed in [17]. In some cases two different slopes may be observed (see for example an application to climatic data [49]). This phenomenon is called a *knee* and at least two explainations have been suggested; it can result from two different types of motion [22] or it can be due to strong correlations between temporally close points which increase the number of near neighbors at small scales and results in an overestimate of the slope [51].

Generally, the error in dimension estimates scales as $O(1/\sqrt{N})$ [91]. However, the number of points needed in a given application has been a major topic of debate. Smith [83] imposed a stringent condition requiring as many as $42^m$ data points, but that was later discarded as far too conservative [93]. On the other hand, it has been claimed that reliable estimates can be obtained even with limited data sets [76]. A formula has been suggested by [69] that gives a minimum number of required points; e.g. for $d \leq 4$ at least 4000 points are required, which is in accordance with many published results. The question of a sufficient $N$ is still subject to study: estimates from small $N$ should not be immediately disregarded although a largest possible $N$ is obviously desirable. Overall, dimension estimates from real data must be considered with caution, particularly when the data are few and noisy. Several published dimension estimates from real data are given in Table (1). Following the discussion above, these estimates could be reliable except in the case of climatology and possibly also of biology and astrophysics.

| field | test | dimension | largest $\lambda$ |
|---|---|---|---|
| electrophysics | electrical circuits with 2 coupled diodes | $N = 40400, \nu \sim 2.5$ (1) | |
| fluid dynamics | low fluid turbulence in Rayleigh-Bernard convection | $N = 20400, \nu \sim 3.0$ (1) | $N = 40000, \lambda \sim 0.01$ (2) |
| fluid dynamics | low fluid turbulence in Couette-Taylor experiment | $N = 16384, \nu \sim 2.4$ (3) | |
| astrophysics | Vela pulsar timing residuals | $N = 564, \nu \sim 1.5$ (4) | |
| climatology | paleoclimatic records: marine-sendiment cores based on 184 raw data | $N = 500, \nu \sim 3.1$ (5) | |
| | | $N = 230, \nu \sim 4.8$ (6) | |
| | | $N = 230, \nu \sim 5.8$ (7) | |
| medicine | ECG cardiac rythm | $N = 60000, \nu \sim 3.6$ (8) | |
| medicine | EEG, short-lived seizure | $N = 6000, \nu \sim 2.05$ (9) | $\lambda \sim 2.9$ (9) |
| medicine | EEG, long-lived epilepsy | $N = 18000, \nu \sim 5.6$ (10) | $\lambda \sim 1.0$ (10) |
| biology | marine plakton diatoms | $N = 830, \nu \sim 2$ (11) | |

Table 1: Estimates of correlation dimension $\nu$ and the largest Lyapunov exponent $\lambda$ from real data of size $N$. The references in the table are: (1): [16], (2): [78], (3): [9], (4): [46], (5): [70], (6): [37], (7): [95], (8): [19], (9): [5], (10): [32], (11): [86].

# 5 Lyapunov exponents

If one wants to study more thoroughly the dynamics of a chaotic system, the Lyapunov exponents should be estimated. The concept of Lyapunov exponents existed long before the establishment of chaos theory, and was developed to characterize the stability of linear as well as non-linear systems. Lyapunov exponents are defined as the logarithms of the absolute value of the eigenvalues of the linearized dynamics averaged over the attractor. The definition covers both discrete and continuous systems[5] [93]. A negative exponent indicates an average rate of contraction for a stable system while a positive value indicates an average degree of expansion of an unstable system. Since the advent of chaos, the set or spectrum of Lyapunov exponents has been considered a measure of the effect of perturbing the initial conditions of a dynamical system. This new role of Lyapunov exponents is not inconsistent with their traditional interpretation. The additional notion is that positive and negative Lyapunov exponents can coexist in a dissipative system, which is then chaotic.

Since the Lyapunov exponents are defined as asymptotic average rates, they are independent of the initial conditions, and hence the choice of trajectory, and therefore they do comprise an invariant measure of the dynamical system. In fact, if one manages to derive the whole spectrum of Lyapunov exponents, other invariants of the system, i.e. the **Kolmogorov entropy** and the attractor's dimension can be found. The Kolmogorov entropy measures the average rate at which information about the state is lost with time. An estimate of this measure is the sum of the positive Lyapunov exponents [74]. The estimate of the dimension of the attractor is provided by the Kaplan and Yorke conjecture [35]:

$$D_L = j + \frac{\sum_{\alpha=1}^{j} \lambda_\alpha}{|\lambda_{j+1}|} \quad (11)$$

where $j$ is defined as $\sum_{\alpha=1}^{j} \lambda_\alpha > 0$ and $\sum_{\alpha=1}^{j+1} \lambda_\alpha < 0$, and the Lyapunov exponents are taken in descending order. $D_L$ gives values close to the dimension estimates discussed earlier and is preferable when estimating high dimensions.

To our knowledge there are two main approaches to computing the Lyapunov exponents. One approach computes the whole spectrum and is based on the Jacobi matrix $Df$ of the system function $f$. The other method computes the largest one or two exponents based on the principal axes of expansion of the system dynamics.

## 5.1 Method based on the Jacobi matrix

Suppose that the dynamics are described by the discrete system $\underline{s}_{k+1} = f(\underline{s}_k)$. If the system is continuous, we work with its discretized version. A small perturbation $\underline{z}_k$ of the trajectory $\underline{s}_k$ is described by the linearized system $\underline{z}_{k+1} = Df(\underline{s}_k)\underline{z}_k$ and after $L$ steps $\underline{z}_{k+L} = Df^L(\underline{s}_k)\underline{z}_k$ where $Df^L(\underline{s}_k) = \prod_{i=0}^{L-1} Df(\underline{s}_{k+i})$. The logarithms of the absolute values of the eigenvalues of this matrix divided by $L$ are first estimates of the Lyapunov exponents, i.e. they are the **local Lyapunov exponents** [24], [2]. They depend on the trajectory $\underline{s}_k$ and show the average divergence from it after $L$ iterations.

The generalization to the invariant **global Lyapunov exponents** is carried out using the Multiplicative Ergodic Theorem of Oseledec [72]. The limit of the matrix $OSL(L,\underline{s}) = \left\{ Df^L(\underline{s}(n))[Df^L(\underline{s}(n))]^T \right\}^{1/2L}$ for $L \to \infty$ exists for a chaotic system and the eigenvalues are independent of the trajectory $\underline{s}$, when $\underline{s}$ is on the attractor [7]. The logarithms of the absolute values of these eigenvalues are then the global Lyapunov exponents.

In practice, for a given system function the Oseledec matrix has to be computed for sufficiently large values of $L$ to assure convergence. Then it becomes ill-conditioned because each eigenvalue is multiplied by $L$ and in a chaotic system the difference between eigenvalues with positive and negative exponents increases exponentially with time. The standard QR-algorithm does not work and a recursive version has been suggested that does not factorize the compound matrix but factorizes the Jacobian matrix step by step [24].

---

[5]For continuous systems the dynamics are defined by the solution of the differential equations, i.e. by the flow.

In the case where only observations are given and the system function is unknown, the matrix $Df$ has to be estimated from the data. In this case, all the suggested methods approximate $Df$ by fitting a local map to a sufficient number of nearby points. The fitted map is typically linear [78], but maps with higher order polynomials seem to offer advantages [15], [11] [6]. Recently, more sophisticated methods based on radial basis functions and neural networks have been applied [25], [67].

When computing Lyapunov exponents from data, an embedding dimension $m$ must be estimated, typically using one of the methods described above. If $m$ is large, some spurious exponents will be computed which will tend to be influenced by noise. To identify these exponents, it has been suggested to repeatedly corrupt the data with artificial noise and rerun the computations. The most sensitive exponents are likely to be the spurious ones [2].

## 5.2 Method based on principal axes of expansion

As shown in Fig(5), a small initial sphere in $m$ space, will turn into an ellipsoide due to contraction and expansion. One may estimate the largest Lyapunov exponents by monitoring the growth of the largest principal axes. In practice, instead of monitoring the evolution of a sphere one follows the evolution of two very close trajectories [7]. Since the divergence of the trajectories are monitored over short time steps one can use the linearized system instead. If the dynamics are unknown, the linear approximation has to be computed from the data as in the first method.

There is an obvious problem in applying this algorithm to chaotic systems because the principal axes quickly become very large and the real divergence of nearby trajectories is hard to follow. To limit the magnitude of divergence, the principal axes must be renormalized according to the Gram-Schmidt renormalization scheme. This is problematic if the attractor is not covered densely enough with points, but an efficient algorithm exists that may help to overcome this drawback [98]. One of the trajectories is always retained for reference. When the vector separating the two trajectories becomes sufficiently large, another data point closer to the reference trajectory is chosen. In this way a few of the largest Lyapunov exponents may be estimated.

Both methods for computing Lyapunov exponents depend critically on the number of data. It has been claimed that the number of points needed for either of the two techniques is about the square of that required to estimate the dimension [23]. On the other hand, published results from simulated data have shown reliable estimates with moderate amounts of data in the order of 1000. In any case, more data are needed to estimate Lyapunov exponents than dimensions. Thus it may not be surprising that Lyapunov exponents have only been estimated from real data in a limited number of cases. This is in contrast to the numerous papers on dimensions from real data. The three estimates of Lyapunov exponents given in Table (1) should be considered with caution although they are derived from comparatively large data sets.

# 6 Conclusion

The initial stimulus to the work reviewed here was to identify deterministic chaos in time series that at first seemed stochastic. The estimation of an attractor's dimension and Lyapunov exponents not only provide an indication of chaos, but also give useful information on the properties of the underlying system, such as the degrees of freedom and level of complexity. However, the reliability of this information is sensitive to the quality and quantity of the data. The need of long time series can hardly be overestimated, particularly if the Lyapunov exponents are desired. In fields like climatology, epidemiology and economy where annual data are explored, better methods relying on less data are generally needed before the presence of chaos can be established with any certainty.

---

[6] In this case a non-linear map is fitted and the linear part is taken as the estimate of $Df$.

Over the last decade, much effort has been devoted to estimate dimensions of strange attractors. This is in contrast to the work on Lyapunov exponents where much less work has been expended. Given the large amount of data needed to compute the Lyapunov exponents, this is not surprising. However, the Lyapunov exponents offer more insight into the dynamics of a chaotic system. Also, their algorithms may be potentially useful in separating chaos from noise thereby providing an independent diagnostic tool for chaotic behaviour. Future work should therefore be directed towards developing algorithms for Lyapunov exponents that are less data intensive.

At present, potentially chaotic behaviour is explored in many different fields. But given the circumstances above, one should be somewhat careful in assessing the results. However, all modellers working with non-linear systems should have some knowledge of the techniques reviewed in this paper. This is necessary since chaos can appear in diverse types of non-linear sytems for certain parameter values. It is useful to be able to diagnose this.

**Acknowledgements** Support for this research was provided by the Norwegian Research Council (NFR) for D. Kugiumtzis and B. Lillekjendlie. B. Lillekjendlie was also supported by SINTEF-SI.

The authors would also like to thank Tom Kavli and Eirik Weyer, both at SINTEF-SI, for valuable discussions and careful reading of the manuscript.

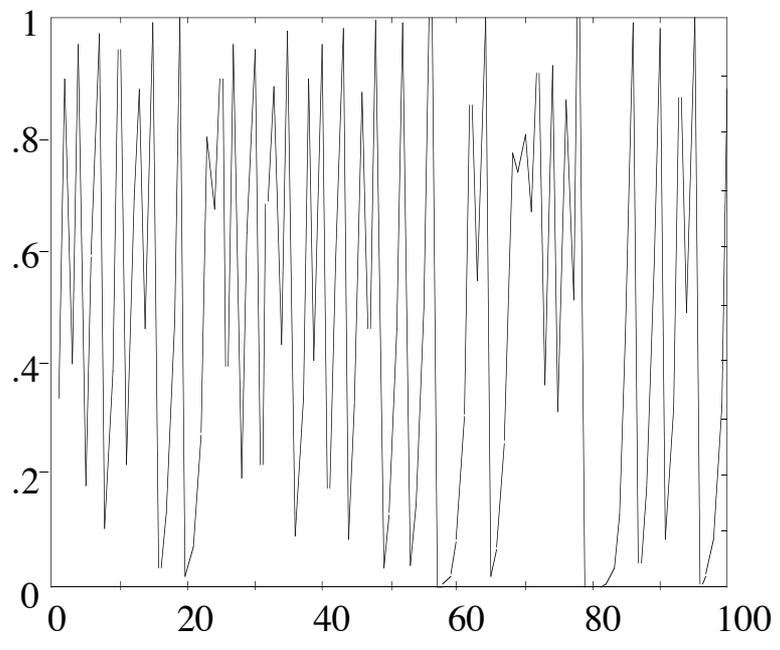

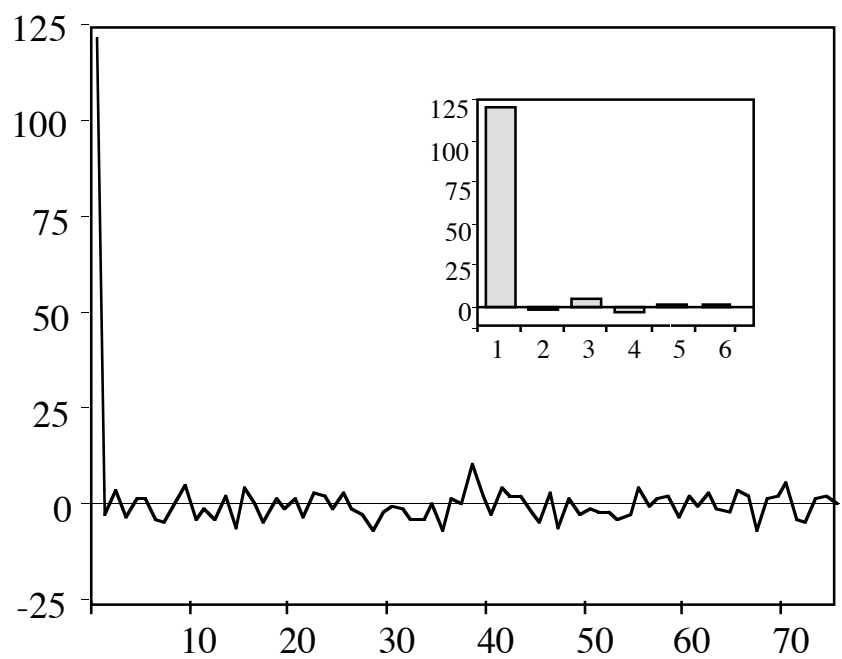

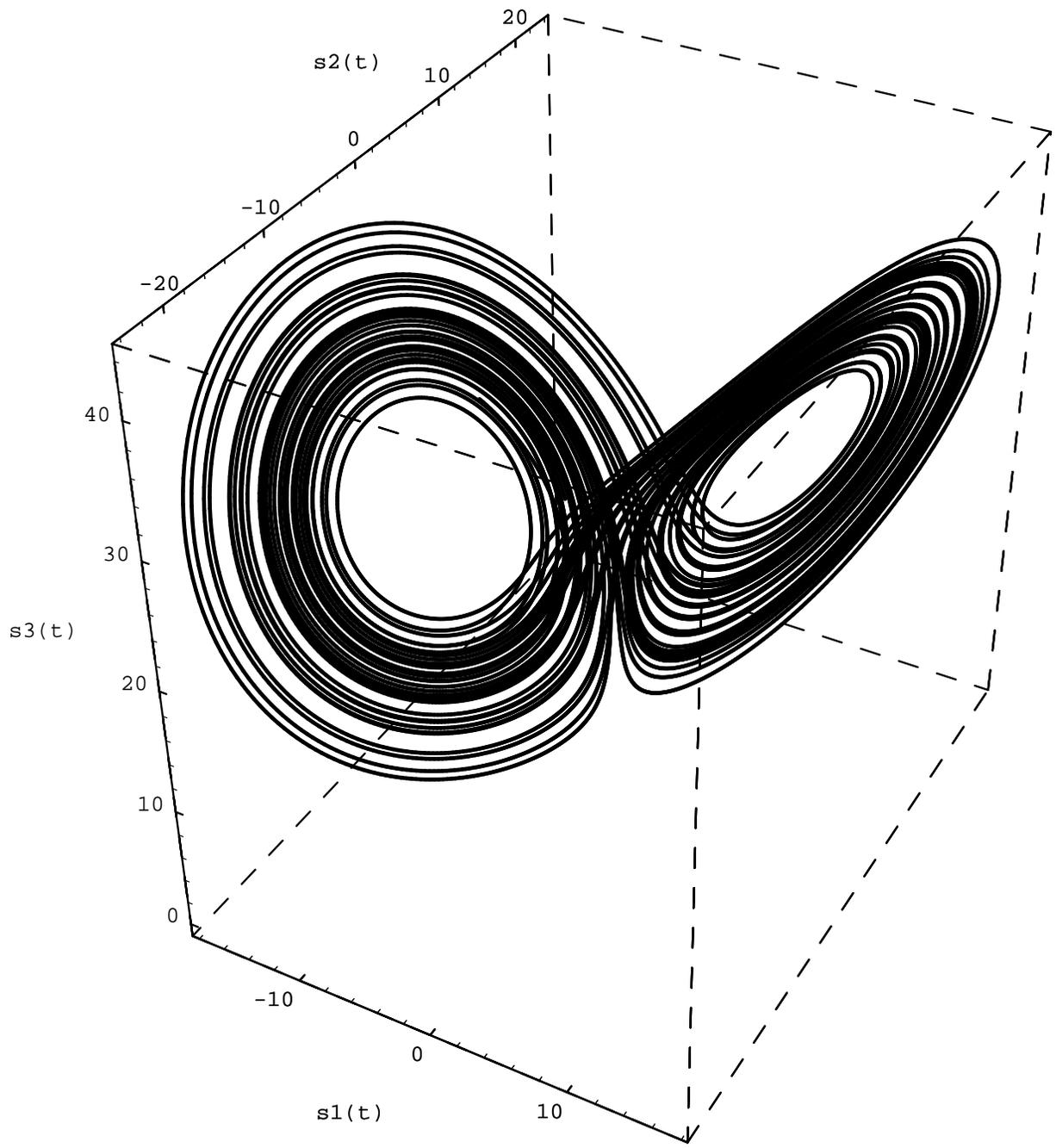

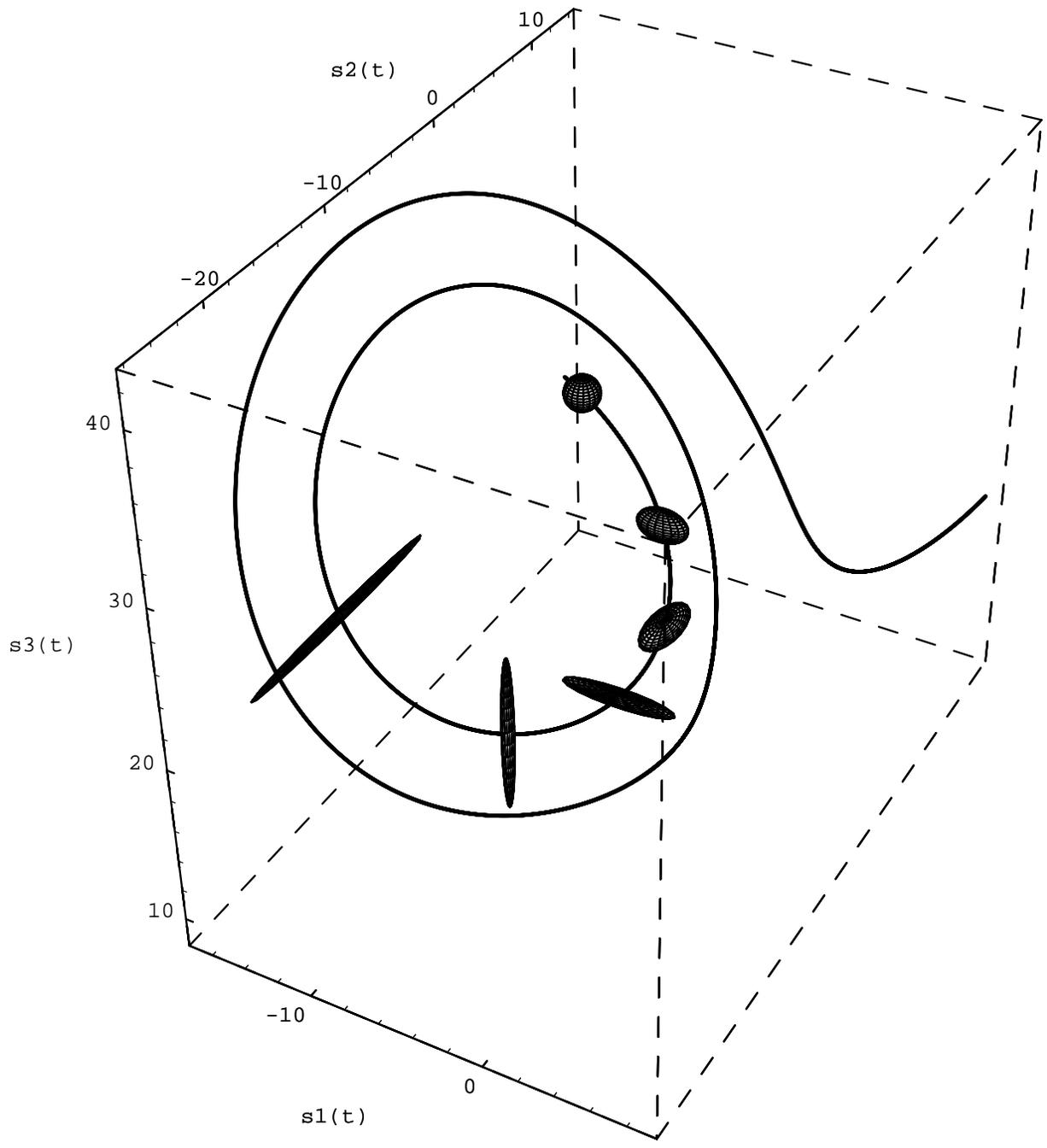

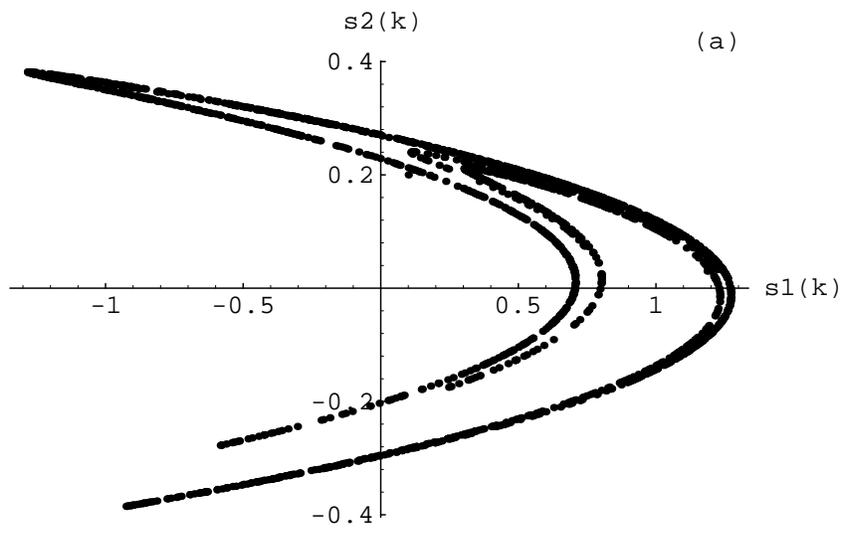
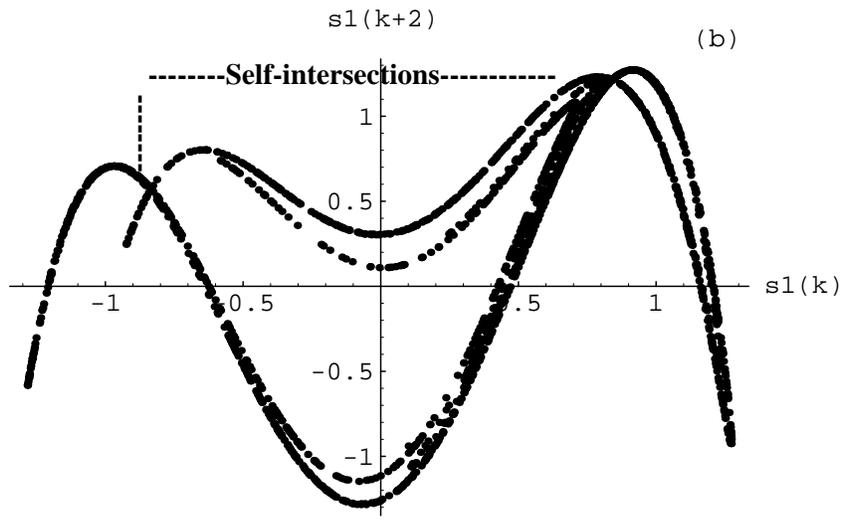
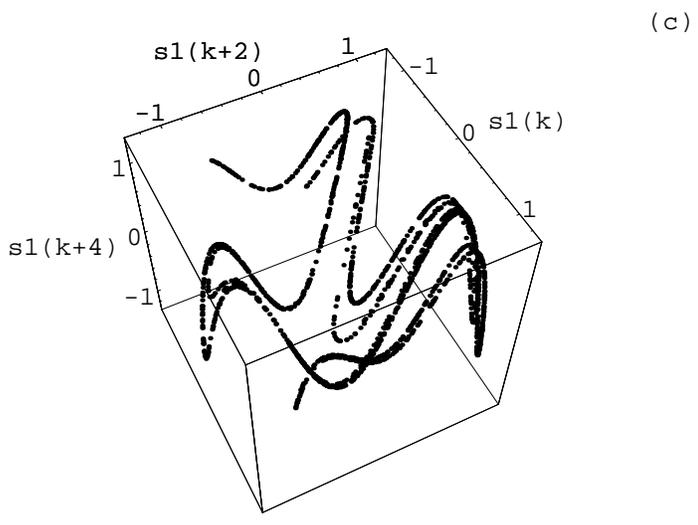

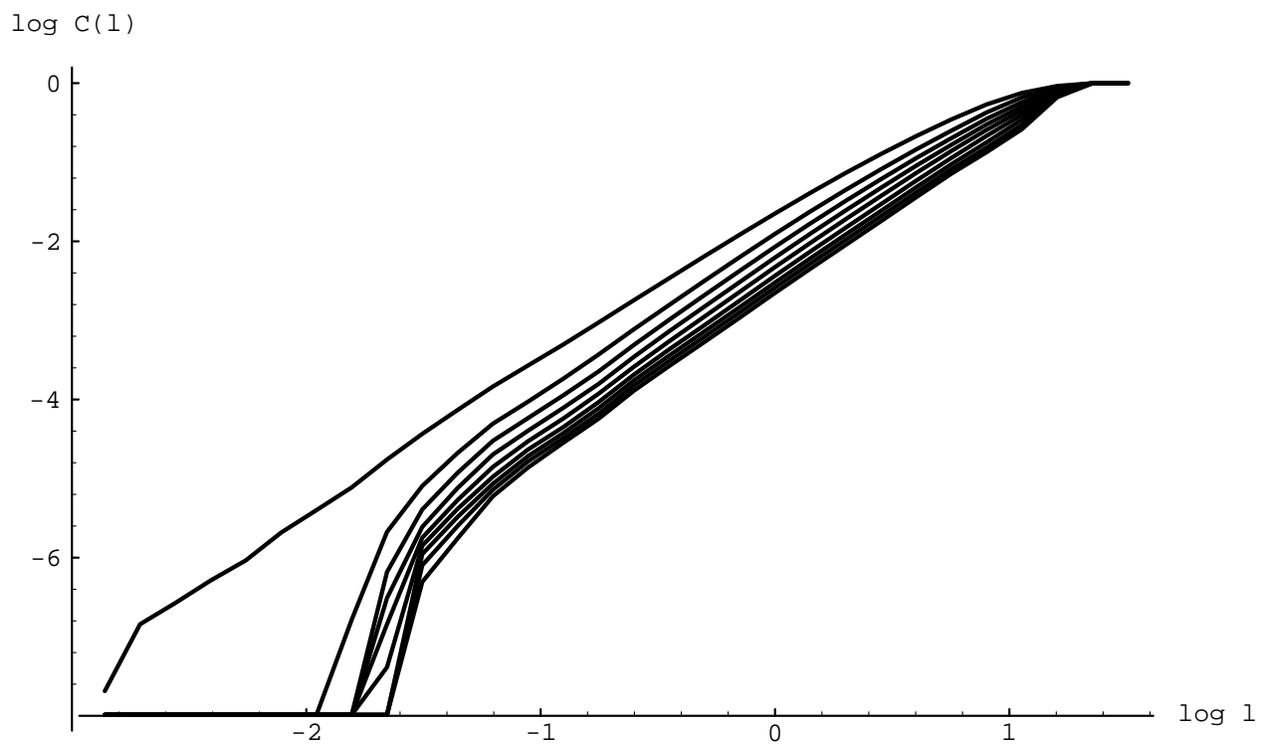

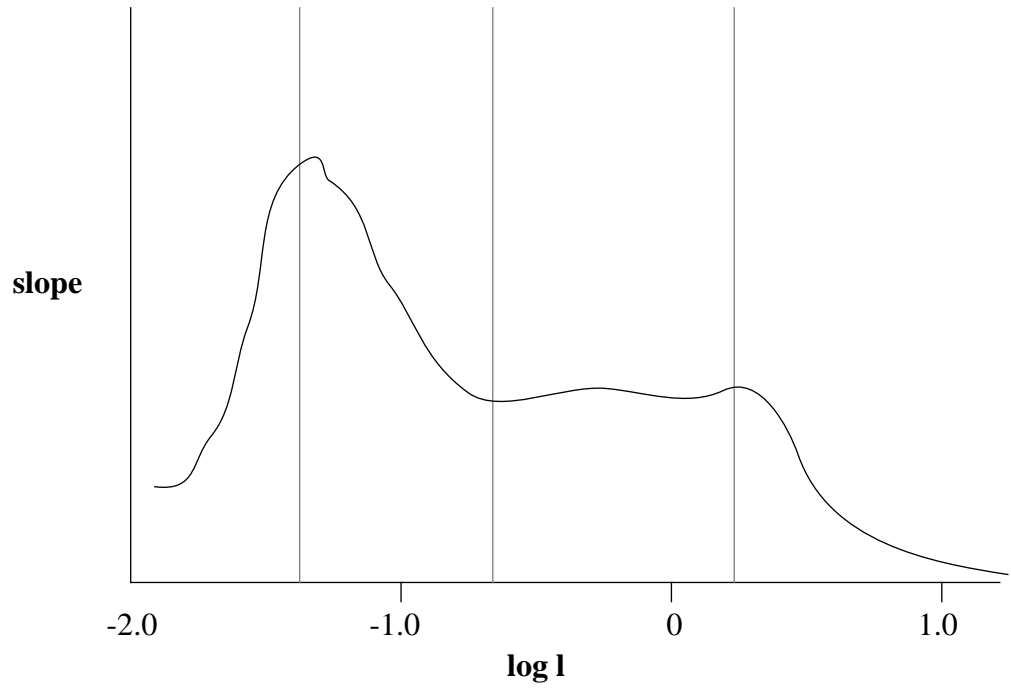